\documentclass[a4paper,fleqn]{cas-sc}

\usepackage[numbers]{natbib}
\usepackage[scientific-notation=true]{siunitx}
\usepackage{amsmath}
\usepackage{braket}
\usepackage{calrsfs}
\DeclareMathAlphabet{\pazocal}{OMS}{zplm}{m}{n}
\newcommand{\Lb}{\pazocal{L}}
\usepackage{tabstackengine}
\stackMath

\def\tsc#1{\csdef{#1}{\textsc{\lowercase{#1}}\xspace}}
\tsc{WGM}
\tsc{QE}
\tsc{EP}
\tsc{PMS}
\tsc{BEC}
\tsc{DE}

\begin{document}
\let\WriteBookmarks\relax
\def\floatpagepagefraction{1}
\def\textpagefraction{.001}
\shorttitle{Round-to-flat and flat-to-round beam transformation}
\shortauthors{S. Kim et~al.}
            
\title [mode = title]{Experimental demonstration of cascaded round-to-flat and flat-to-round beam transformations}                   
\author[1]{Seongyeol Kim}[orcid=0000-0003-1431-4732]
\cormark[1]
\ead{sykim12@postech.ac.kr}

\credit{Design of this study, Methodology, Experiment, Acquisition of data, Data analysis, Interpretation, Validation, Drafting, Reviewing, Revising}

\affiliation[1]{organization={Argonne Wakefield Accelerator Group},
                addressline={Argonne National Laboratory}, 
                city={Lemont},
                postcode={60439}, 
                state={IL},
                country={USA}}

\author[1]{Philippe Piot}[orcid=0000-0002-4799-292X]

\credit{Conceptualization and design of this study, Methodology, Experiment, Interpretation, Validation, Supervision, Drafting, Reviewing, Revising, Final Approval}


\author[1]{Gongxiaohui Chen}
\credit{Experiment, Validation, Reviewing}

\author[1]{Scott Doran}
\credit{Experiment, Validation, Reviewing}

\author[1]{Wanming Liu}
\credit{Experiment, Software, Validation, Reviewing}

\author[1]{Charles Whiteford}
\credit{Experiment, Validation}

\author[1]{Eric Wisniewski}
\credit{Experiment, Validation, Reviewing}

\author[1]{John Power}
\credit{Conceptualization and design of this study, Methodology, Drafting, Validation, Supervision, Reviewing, Revising, Final Approval}

\cortext[cor1]{Corresponding author. Present address: Pohang Accelerator Laboratory, POSTECH, Republic of Korea.}

\begin{abstract}
Magnetized beams $-$ beam with significant canonical angular momentum $-$ are critical to electron cooling of hadron beams such as contemplated in next-generation hadron and electron-ion colliders. The transport of magnetized electron beams over long distances in a locally non-axisymmetric external field is challenging. An alternative is to transform the beam into an uncoupled "flat beam", transport the produced "flat" beam over a long distance, and reintroduce the cross-plane coupling to "re-magnetize" the beam. In this paper, we demonstrate via numerical simulation and laboratory experiments such a cascaded-transformation approach. 
\end{abstract}

\begin{keywords}
Electron cooling \sep Magnetization \sep Beam transformation 
\end{keywords}

\maketitle

\section{Introduction}

Stored high-intensity beams have been ubiquitous tools in elementary particle physics to support discovery science in high-energy physics but also precise measurement of cross-sections for processes in quantum electrodynamics. 
An important parameter quantifying the capabilities of these beams  $-$ the luminosity $-$ is ultimately limited by the phase-space density attainable in the storage rings~\cite{LEHRACH2006289}. 
For hadron beams effects such as beam-beam interaction during the collision, intrabeam effects (such as scattering and possibly space charge), along with scattering on residual gas set the maximum luminosity that can be reached. 
Therefore, phase-space cooling $-$ a method to reduce the phase space volume $-$ has been implemented in hadron rings and is being considered for most of the next-generation hadron colliders under design~\cite{blaskiewicz2014cooling}. 
Consequently, devising efficient and fast phase-space cooling techniques is critical for future hadron and electron-ion colliders~\cite{Poth-Nature1990,VanDerMeer-RevModPhys.57.689, Mikhailichenko-PhysRevLett.71.4146, Zolotorev-PhysRevE.50.3087, PhysRevLett.61.826}.

In electron cooling~\cite{Budker:1967sd, Budker:1021068}, a bright electron beam co-propagates with the ion beam. In the ion-beam rest frame, the electrons (which have the same average velocity) produce an equivalent friction force which reduces the transverse velocity of the electron. 
Since its experimental demonstration fifty years ago~\cite{Budker:1967sd, Budker:1021068}, electron cooling has been implemented in several proton and ion storage rings. In most configurations, the electron and hadron beams interact within a long solenoid providing a uniform axial magnetic field $B_s$. 
In addition to guiding the electron beam, the magnetic field can be combined with an angular-momentum-dominated -- or magnetized -- electron beam to possibly enhance the cooling efficiency~\cite{Derbenev:1978qd,Derbenev:1978,Derbenev2000AdvancedOC}. Given the conservation of magnetic flux, the electrons do not have any ``coherent" motion but instead, all electrons cycle on small helices with a radius given by the gyroradius. Producing, accelerating, and transporting relativistic (Lorentz factor $\gamma \ge 50$) magnetized beams $-$ i.e. beams with large kinetic angular momentum $-$ with parameters consistent with future electron-ion colliders~\cite{deitrick24development, willeke2021electron,Montag:2019rpr, Zhang:2019afy, Benson:2018jzt} is a challenging task. The formation of magnetized beams in photoinjectors and their conversation to flat beam has been demonstrated in several conceptual studies and experiments~\cite{brinkmann-2001-a, Sun:PhysRevSTAB.7.123501, PhysRevSTAB.9.031001, Halavanau:2018hcg, PhysRevSTAB.4.053501, Xu:2019xjg, PhysRevAccelBeams.25.044001, WIJETHUNGA2023168194}.

In proposed hadron-cooler designs using magnetized beams, the generated magnetized beam is transformed into a flat beam, for acceleration and transport to the cooling insertion where it is transformed back into a magnetized beam. The transformation from round to flat and flat to round magnetized beam use skew-quadrupole magnets. 
To explore such a strategy, we designed and implemented an experiment at the Argonne Wakefield Accelerator (AWA) facility. 
In the experiment, a magnetized beam originating from a photoinjector is transformed to a flat beam by a Round to Flat Beam Transform (RFBT), transported $\sim 10$~m, and then transformed back to a magnetized beam by a Flat to Round Beam Transform (FRBT). 

In the following section, we briefly introduce to the phase-space-manipulation concept underlying this study.
In Section~\ref{numerical_simulations}, we present simulation results and discussions related to the experimental setup.
Finally, we present the experimental demonstration of the beam transformations at the AWA facility in Sec.~\ref{expdemo}.

\section{Phase-space repartitioning in 4D}\label{Intro}
This section briefly introduces the underlying concepts necessary to understand the phase-space manipulations investigated thoughout this paper. 
We consider the 4D phase space $\mathbf{X, Y}^T$ with $\mathbf{X}=(x,x')^T$ and $\mathbf{Y}=(y,y')^T$ ($^T$ representing the transposition operator) where $x$ (resp. $y$) represents the transverse horizontal (resp. vertical) position associated with an electron and $x'\equiv p_x/p_z$ (resp. $y'\equiv p_y/p_z$) the horizontal (resp. vertical) angle of its trajectory to the longitudinal axis $\hat z$ [here $p_i$ ($i=[x,y,z]$) is the momentum along the given direction]. 

The angular-momentum-dominated (magnetized) beam is initially produced in an RF photocathode gun~\cite{Sun:PhysRevSTAB.7.123501}. The RF gun is nested in a pair of solenoidal lenses tuned to produce a non-vanishing axial magnetic field ${\mathbf B}=B_{c}\hat z$ at the photocathode surface. 

To macroscopically characterize the impact of angular momentum on a beam, it is convenient to introduce the magnetization 
\begin{equation}
\Lb = \frac{\braket{L_{z}}}{2m_{e}c} = \frac{eB_{c}\sigma_{c}^{2}}{2m_{e}c}.
\label{eq:magnetization}
\end{equation}
$\Lb$ is an averaged value of the angular momentum over the beam distribution. 
Here ${L_{z}}=\frac{eB_{c}}{2}(x^{2}+y^{2})$ represents the angular momentum generated by non-zero magnetic field at the cathode, $\sigma_{c}$ is the RMS size of the laser on the photocathode, $m_{e}$ is the electron mass, and $c$ is the speed of light. 

To describe the property of the magnetized beam we introduce the beam's $4\times 4$ covariance matrix as 
\begin{equation}
\Sigma = \left[\begin{array}{cc} 
\braket{\textbf{X}\textbf{X}^T} & \braket{\textbf{X}\textbf{Y}^T} \\
\braket{\textbf{Y}\textbf{X}^T} & \braket{\textbf{Y}\textbf{Y}^T} \\
\end{array}\right].
\label{eq:sigmamatrix}
\end{equation}
For a non-vanishing kinetic angular momentum, the off-diagonal terms are given by
\begin{equation}
\braket{\textbf{X}\textbf{Y}^T} = -\braket{\textbf{Y}\textbf{X}^T} = \Lb J,
\label{eq:sigma_relation_magnetization}
\end{equation}
where $J \equiv \left[ \begin{array} { c c } 0 & 1 \\ - 1 & 0 \end{array} \right]$. 

Here, through the diagonalized $\Sigma$ matrix by symplectic transformation and invariance of 4D emittance, we can obtain the eigenemittance~\cite{Burov-PhysRevE.66.016503,Kim:PhysRevSTAB.6.104002, PhysRevAccelBeams.24.054201}.
\begin{equation}
\epsilon_{\pm} = \sqrt{\epsilon_{u}^{2} + \Lb^{2}} \pm \Lb,
\label{eq:eigenemittances}
\end{equation}
where $\epsilon_{u}$ is the uncorrelated emittance. 

These eigenemittances can be mapped to the two projected transverse emittances when a decoupling transformation is applied~\cite{Kim:PhysRevSTAB.6.104002}.
When the magnetization is much larger than the uncorrelated emittance, then the eigenemittances become
\begin{equation}
\epsilon_{+} \simeq 2\Lb, ~~ \epsilon_{-} \simeq \frac{\epsilon_{u}^{2}}{2\Lb}.
\label{eq:eigenemittances_simplified}
\end{equation}
Thus, it is expected that the emittances in horizontal and vertical plane becomes asymmetric when the angular momentum is dominated on the beam.
We note that, for a case where the beam is completely decoupled, the eigenemittance becomes normalized emittance with subscript "n."
The beam with large emittance ratio ($\epsilon_{+}/\epsilon_{-} >> 1$) is called flat beam.

The decoupling of the beam can be implemented using three skew-quadrupole magnets~\cite{brinkmann-2001-a}. The decoupling process in the RFBT can be conveniently described using a correlation matrix $C$ defined as $\mathbf{Y}=C\mathbf{X}$~\cite{thrane2002photoinjector}. For a magnetized beam, this correlation matrix can be written as~\cite{PhysRevAccelBeams.25.044001}
\begin{equation}
C = \frac{\Lb}{\epsilon_{eff}}\left[\begin{array}{cc} 
\alpha_{RF} & \beta_{RF} \\
-\frac{1+\alpha_{RF}^{2}}{\beta_{RF}^{2}} & -\alpha_{RF} \\
\end{array}\right].
\label{eq:Cmatrix}
\end{equation}
Here, $\beta_{RF}$ and $\alpha_{RF}$ are the Twiss parameters associated with the incoming magnetized beam at the first skew-quadrupole magnet entrance.
A subscript $RF$ indicates the beam parameter are given at the entrance of the RFBT.
Note that $\beta_{x}^{RF}(\alpha_{x}^{RF})$ = $\beta_{y}^{RF}(\alpha_{y}^{RF})$.
When $\Lb\gg\epsilon_{u}$, the prefactor in Eq.\eqref{eq:Cmatrix} is  $\frac{\Lb}{\epsilon_{eff}}\simeq 1$.
Therefore, using Eq.\eqref{eq:Cmatrix} together with the skew matrix, one can obtain the solution for optimized skew quadrupole strengths given the incoming beam's Twiss parameters.

In case of FRBT process, detailed descriptions of how to obtain the required parameters and skew strengths are reported in Ref.~\cite{PhysRevSTAB.9.024001}.
The obtained solution which assumes a vanishing divergence for the incoming beam (i.e Twiss parameter $\alpha_{x,y}^{FR}=0$ at the entrance of FRBT section) is
\begin{eqnarray}
\beta_{x,y}^{FR} &=& 2L_{d}\sqrt{1+\sqrt{2}},\nonumber\\
k_{2} &=& \frac{2\sqrt{2}}{\beta_{x,y}^{FR} L_{q}}, \mbox{~and}\\ 
k_{1} &=& k_{3} = -\frac{1}{2}\left(1+\frac{1}{\sqrt{2}}\right)k_{2},\nonumber
\label{eq:FRBT_eq}
\end{eqnarray}
where $\beta_{x,y}^{FR}$ is the Twiss betatron function of the incoming flat beam at the beginning of the FRBT section, and $L_{d}$ is a drift space between skew quadrupole magnets, and $L_{q}$ is a magnetic length of the skew-quadrupole magnet.
The superscript $FR$ indicates the FRBT.
$k_{1,2,3}$ are the normalized quadrupole-magnet strengths in units of~\si{\meter}$^{-2}$. 
Although the matching conditions described by Eq.~\eqref{eq:FRBT_eq} are limited to the case of $\alpha_{FB}=0$, work on developing a more general formalism is challenging and alternative technique for the more general cases rely on numerical method, e.g. the use adjoint technique-based FRBT optimization as discussed in Ref.~\cite{PhysRevAccelBeams.25.044002}. In this paper we extensively rely on numerical simulation and a complete study of the cascaded round-to-flat and flat-to-round beam transformations performed with particle-tracking simulations is discussed in the next section.

\begin{figure*}[t!]
   \centering
   \includegraphics[width=\textwidth]{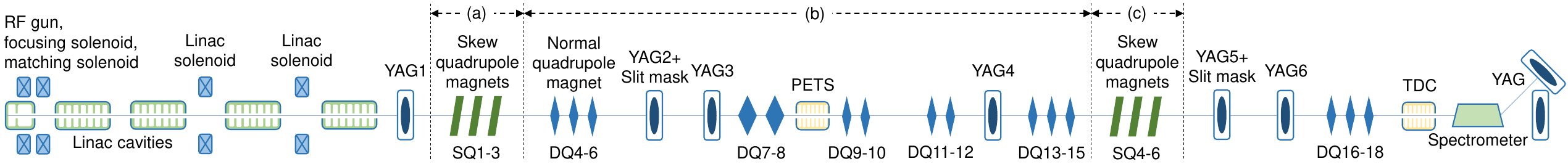}
   \caption{AWA drive linac and beam transport line (not to scale)~\cite{kim:ipac2023-wepa037}. 
   The AWA linac has six accelerating cavities, but only four cavities were used in the present experiment.
   (a): First skew-quadrupole magnets for the RFBT. 
   (b): Flat-beam transport line. Here, 12 normal quadrupole magnets were mainly used for flat beam transport.
   (c): Second skew-quadrupole magnets for the FRBT.}
   \label{fig:AWASchematic}
\end{figure*}

\section{Particle tracking simulations}\label{numerical_simulations}

We performed start-to-end particle tracking simulations using OPAL~\cite{https://doi.org/10.48550/arxiv.1905.06654} particle-tracking code to understand the coupled beam dynamics during round-to-flat and flat-to-round beam transformations and to have a reference setup for RFBT, beam transport, and FRBT along the AWA beamline. 
The OPAL code includes a model of the photoemission process and consider space-charge effects using a quasi-static approach. 
A diagram of the AWA beamline appears in Fig.~\ref{fig:AWASchematic}. The beamline consists of a photoinjector, RFBT skew-quadrupole-magnets, an uncoupled transport line (UTL), and FRBT skew-quadrupole-magnets. 
Three solenoidal lenses are used to control the magnetic field in the RF gun area a pair of identical solenoids (the bucking and focusing solenoids) are symmetrically located to nominally zero the magnetic field on the photocathode surface.
In our configuration, the bucking solenoid was turned off to allow for a tunable magnetic field $B_c\le 0.14$~T on the photocathode controlled by the focusing solenoid.
Another solenoidal lens $-$ the matching solenoid $-$ is employed to control the magnetized-beam emittance.

Table~1 summarizes the accelerator settings used in the tracking simulations.
\begin{table}[b]
\label{table111} 
\caption{
Photocathode UV-laser pulse parameters and accelerator beamline settings used in the simulations \label{tab:tab1}.
}
\begin{tabular*}{\tblwidth}{@{} LR@{} }
\toprule
\textrm{Parameter} & \textrm{Value} \\
\midrule
UV laser parameters &  \\ 
Charge, $Q$ & 1.0~nC \\
UV radius & 2.7~mm  \\
UV pulse length (FWHM) & 3.0~ps \\
\midrule
Photoinjector parameters &  \\ 
Beam energy, $E_{kin}$ & 40~MeV  \\
Gun gradient & 60~MV/m \\
Gun phase, $\phi_{G}$ & $-$25~deg from nominal 50 deg \\
Linac 1-5 gradient & 22~MV/m \\
Linac 1,2 phase, $\phi_{L1,2}$ & $-$12 deg from on-crest \\
Linac 3,5 phase, $\phi_{L3,5}$ & $-$25 deg from on-crest \\
Magnetic field at the cathode & 0.14~T \\
\bottomrule
\end{tabular*}
\end{table}
To mitigate the space-charge-induced energy spread in the low-energy regime, the ultraviolet (UV) laser pulse was configured to be a 3-ps (FWHM) long pulse. 
Such a long pulse also prevents the accumulation of large correlated energy spread typically associated with the use of shorter laser pulses~\cite{PhysRevAccelBeams.25.044001,PhysRevLett.100.244801}. 
The longitudinal distribution is a plateau distribution generated by stacking 16 300-fs (FWHM) UV laser pulses using 4 birefringent $\alpha$-BBO crystals present at the AWA facility. 
Such a stacking technique results in the plateau distribution having rising and falling edges with a duration commensurate with the initial laser duration (300 fs).
The transverse UV-pulse distribution is taken to be radially uniform.
In addition, the RF gun and linac cavity phases were adjusted to cancel the energy chirp $\partial E / \partial z$ in an attempt to minimize chromatic aberrations in the subsequent beamline elements. 
The final beam energy is 40~MeV which is relevant to possible electron-cooling options, e.g., in the electron-ion collider (EIC) in construction at BNL~\cite{deitrick24development, willeke2021electron}.

\begin{figure*}[ht]
   \centering
   \includegraphics[width=\textwidth]{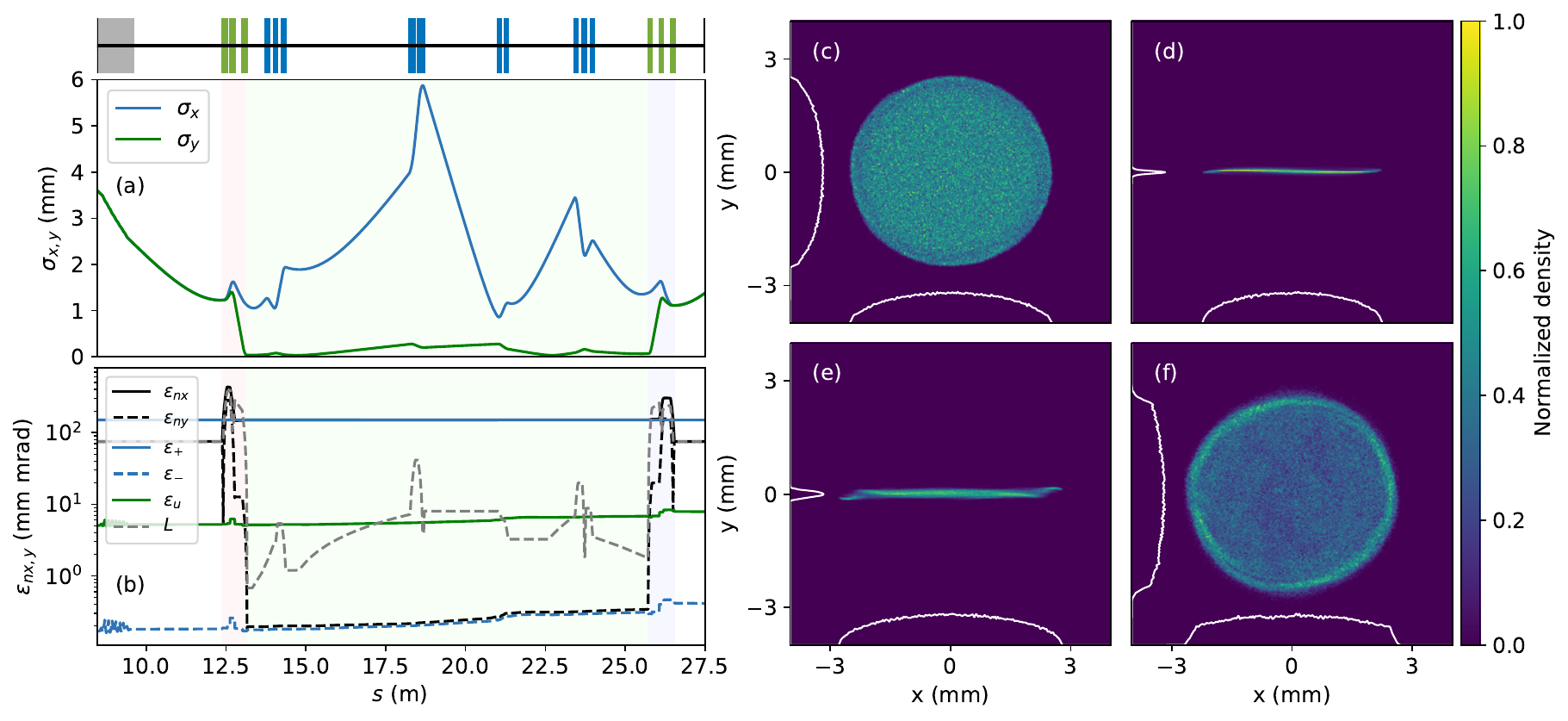}
   \caption{
   (a) RMS beam size and (b) emittances and calculated magnetization along the RFBT, flat-beam transport, and FRBT line. 
   At the beginning, grey block indicates the last accelerating cavity. 
   $(x-y)$ beam distributions are shown on the right side. 
   (c): After AWA drive-linac section. 
   (d): Right after the RFBT section. 
   (e): Right before entering the FRBT section. 
   (f): After FRBT section.}
   \label{fig:opalsim_sigemit_beamdist}
\end{figure*}

Figure~\ref{fig:opalsim_sigemit_beamdist}(a,b) shows the evolution of the RMS beam sizes, emittances, and magnetization along the beamline downstream of the linac cavity.
The simulation considered a magnetic field on the photocathode $B_c=0.14$~T for which the theoretically-predicted magnetization is $\Lb= 74.7$~\si{\micro\meter}.
At the RFBT section around $s=12.5$~m, cylindrical-symmetric ($\sigma_x=\sigma_y$ and $\epsilon_{nx}=\epsilon_{ny}$) beam [see Fig.~\ref{fig:opalsim_sigemit_beamdist}(c)] becomes flat downstream of the RFBT [see Fig.~\ref{fig:opalsim_sigemit_beamdist}(d)] with its transverse emittances mapping to the eigenemittances described in Eq.\eqref{eq:eigenemittances_simplified}.
In the simulations the RFBT condition derived in~\cite{thrane2002photoinjector} is used to devise the skew-quadrupole magnets settings that cancel the $\gamma\langle xy' \rangle$ and $\gamma\langle yx' \rangle$ coupling terms. 
These values are used as a starting point and further optimized in {\sc opal}; the comparison of the numerically-optimized and theoretical values appear in Table~2.
Error between values from analytic formula and simulation is less than 4\%.
\begin{table}[b]
\caption{
Comparison of skew quadrupole strengths from analytic formula and OPAL simulations.
}
\begin{tabular*}{\tblwidth}{@{} LCC@{} }
\toprule
\textrm{Case} & \textrm{Analytic} & \textrm{OPAL} \\
\midrule
RFBT & & \\
SQ1 (T/m) & $+$2.717 & $+$2.704 \\
SQ2 (T/m) & $-$3.501 & $-$3.455 \\
SQ3 (T/m) & $+$3.094 & $+$2.985 \\
\midrule
FRBT & & \\
SQ4 (T/m) & $-$2.675 & $-$2.981 \\
SQ5 (T/m) & $+$3.134 & $+$3.427 \\
SQ6 (T/m) & $-$2.675 & $-$2.623 \\
\bottomrule
\end{tabular*}
\end{table}
\label{table:analytic_opal_rfbtskewst}

Downstream of the RFBT section, the magnetized beam becomes a decoupled flat beam. 
Thus the eigenemittances become projected emittances in each plane.
The projected normalized emittances of the flat beam are $\epsilon_{nx}=149.9$~\textmu{m}, $\epsilon_{ny}=0.2$~\textmu{m}. 
These emittances follows the relations described in Eq.~\eqref{eq:eigenemittances_simplified}.

The flat beam is transported along the UTL section which is a 12.6-meter-long beamline that incorporates normal quadrupole magnets. 
The UTL quadrupole-magnet settings were optimized to minimize the emittance growth due to collective effects and aberrations. 

The magnetization along the flat beam transport section illustrated in Fig.~\ref{fig:opalsim_sigemit_beamdist}(b) was calculated through Eq.~\eqref{eq:sigmamatrix}. 
Then, the magnetization can be re-written as
\begin{equation}
\Lb=\frac{|\gamma\left(\braket{xy'}-\braket{yx'}\right)|}{2}.
\label{eq:Magnetization2}
\end{equation}
Therefore, even through the theoretical $\Lb$ of Eq.\eqref{eq:magnetization} itself is invariant, the calculated $\Lb$ becomes evolving due to the additional contribution of the intrinsic transverse beam parameters.
As can be seen, the magnetization increases by the order of magnitude when the flat beam size becomes large around $s=17.5$~m in Fig.~\ref{fig:opalsim_sigemit_beamdist}(b). 

In addition, during the flat beam transport, we observed that the vertical emittance $\epsilon_{ny}$ (corresponding to the smallest of the eigenemittances) increases to from 0.20 to 0.35~\textmu{m}.
We found that the beam is still space-charge dominated according to Laminar criterion~\cite{ferrario2020injection}.
Therefore, the space charge force affects the fractional energy spread to be large, resulting in the increases of chromatic aberrations.
Nevertheless, the flat-beam right before entering the FRBT section as shown in Fig.~\ref{fig:opalsim_sigemit_beamdist}(e) maintains its flatness.

For the FRBT to convert the beam back to the magnetized state, we utilized the FRBT condition in Eq.\eqref{eq:FRBT_eq}; we optimized the transport line optics to match the Twiss parameters of the beam accordingly. 
First, incoming Twiss $\beta_{x,y}$ was matched to 1.08 m with given drift space between the skew quadrupole magnets.
Whereas, we found that beam divergence still exists; incoming Twiss $\alpha_{x}(\alpha_{y})=-0.2(-0.3)$.
Nevertheless, as listed in Table~2, quadrupole strengths found from the tracking simulation are not significantly deviated compared to those from the analytic condition.

As shown in Fig.~\ref{fig:opalsim_sigemit_beamdist}(a,b), the RMS beam envelope in horizontal and vertical plane becomes identical.
In addition, the projected emittances and magnetization return to  original values before the RFBT block, confirming that the FRBT process is completed.
Likewise, the transformed round beam through the FRBT section is shown in Fig.~\ref{fig:opalsim_sigemit_beamdist}(f).
It implies that the small amount of increases of the uncoupled flat beam emittance due to the chromatic aberration does not play a significant role in distorting the magnetized state of the transformed round beam.

\begin{figure}[ht]
   \centering
   \includegraphics[width=8.3 cm, keepaspectratio]{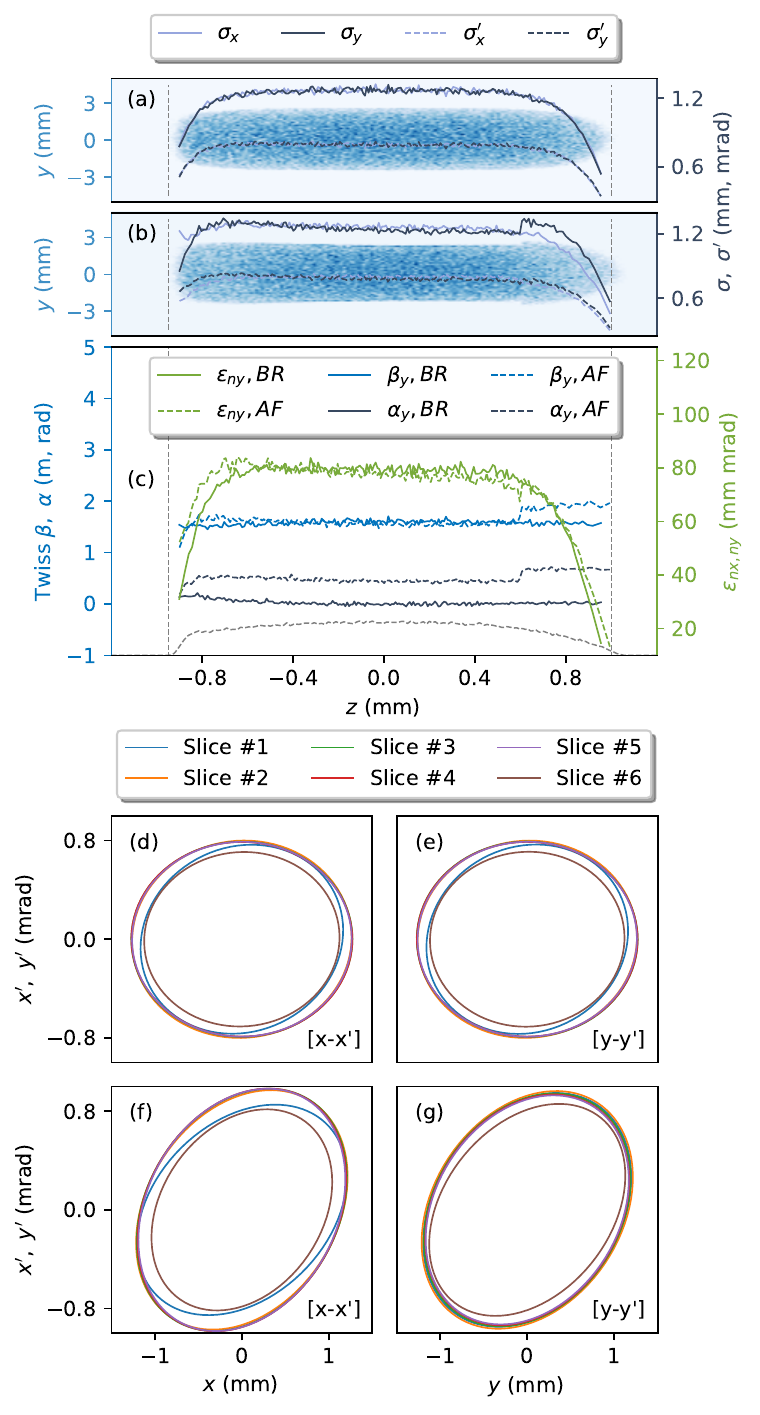}
   \caption{Slice beam distribution (a) before RFBT and (b) after FRBT sections, respectively. 
   Head of the beam is placed at $z>0$, while the tail at $z<0$.
   Plots in (a-b) indicate the RMS envelope and divergence along the slice.
   (c) Vertical Twiss parameters and normalized emittances along the beam slice. 
   Grey dashed line on the ground shows the slice density distribution.
   \textit{BR} indicates "Before RFBT" and \textit{AF} represents "After FRBT."
   On the bottom, slice phase spaces (d-e) before RFBT and (f-g) after FRBT are depicted where the first(last) slice refers to that in the tail(head) part.}
   \label{fig:simulated_slicephasespace}
\end{figure}

It is worth noting that, interestingly, the final round beam exhibits a ring-shape feature. To understand the origin of this feature, we investigated the evolution of temporal slices of the transverse phase spaces.
Figure~\ref{fig:simulated_slicephasespace} illustrates the slice beam phase spaces and parameters.
First, as depicted in Fig.~\ref{fig:simulated_slicephasespace}(a), the slice distribution of the initial round, magnetized beam before entering the RFBT has almost identical beam envelope and divergence except for the head and tail sections.
These correspond to the slice emittance and Twiss parameters illustrated in Fig.~\ref{fig:simulated_slicephasespace}(c) and the resulting slice phase space Fig.~\ref{fig:simulated_slicephasespace}(d,e).
Here, as mentioned, the slice parameters at head and tail sections are different.
It implies that the betatron oscillation at those sections is eventually different compared to that in the center of the beam. 
Therefore, after the FRBT section, the slice beam envelope at the tail $(z<-0.6$~mm) and head $(z>+0.6$~mm) parts becomes larger than those of the initial state, as shown in Fig.~\ref{fig:simulated_slicephasespace}(b).
Similarly, the increases of Twiss betatron function in vertical plane is observed [see Fig.~\ref{fig:simulated_slicephasespace}(c)].

Particularly, this evolution of the slice parameters results in the matching of the slice phase space in vertical plane as illustrated in Fig.~\ref{fig:simulated_slicephasespace}(g) in terms of the size and divergence of the envelope.
Therefore, this phase space is mixed with the horizontal phase space, resulting in the ring shape at the edge of the beam distribution.
Furthermore, this feature is also observed in the case where the space charge effect and chromatic aberration are not considered in the tracking simulation.

The particle tracking simulation confirm that cascading the round-to-flat and flat-to-round beam transformation globally converse the beam parameters and especially the beam magnetization even when considering space-charge effect and a realistic beamline. In addition, the simulations indicate the formation of a ring-shaped feature that appears to originate some slice mismatch along the beam.

\section{Experimental demonstration\label{expdemo}}

\begin{figure}[t]
   \centering
   \includegraphics[width=8.3 cm, keepaspectratio]{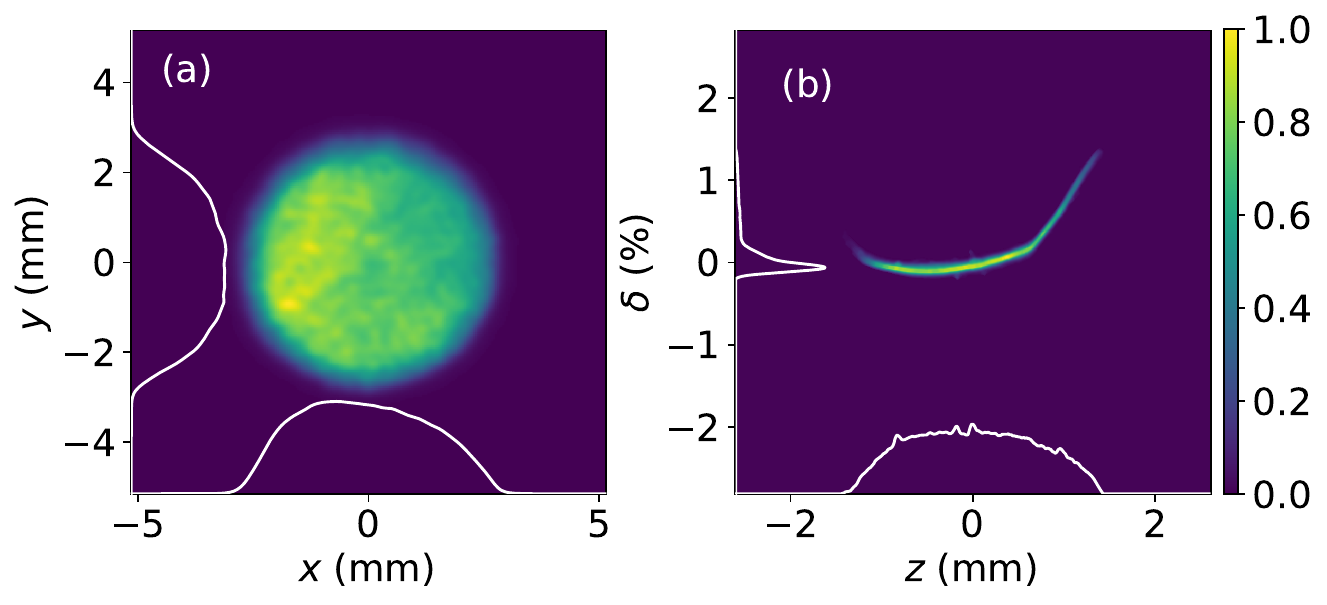}
   \caption{(a) Initial UV distribution measured at the virtual cathode and (b) measured LPS at the end of the straight beamline. 
   Beam head is placed at positive $z$.
   Some spikes on the horizontal histogram of LPS are due to radiated lights that are not removed during the post-processing of the data.}
   \label{fig:UV_LPS}
\end{figure}

The tracking simulation results guided the design of an experiment at AWA. The experimental conditions used for the demonstration experiments are summarized in Table~\ref{tab:tab1}.
Transverse profile of the UV laser illuminating the cathode, as illustrated in Fig.~\ref{fig:UV_LPS}(a), was optimized to be uniform using a micro-lens array~\cite{PhysRevAccelBeams.20.103404}.
Diameter of the transverse UV profile is set to about 5.4~mm.
Particularly, we adjusted the RF cavity phases during the experiment to obtain minimized energy spread.
The measured final beam energy downstream of the linac is $43.4\pm 0.03$~MeV, which is slightly different compared to the simulation case; it is expected that there is systematic error of the RF cavity gradient. Likewise, the RMS relative energy spread is measured to be 0.3\%. The measured longitudinal phase space appears in Fig.~\ref{fig:UV_LPS}(b) and display some distortion with the head of the bunch (corresponding to $z>0$) having a chirp. 
Except for that region, overall the energy chirp is controlled and the RMS energy spread is approximately 0.05\%.

\begin{figure}[b]
   \centering
   \includegraphics[width=8.3 cm, keepaspectratio]{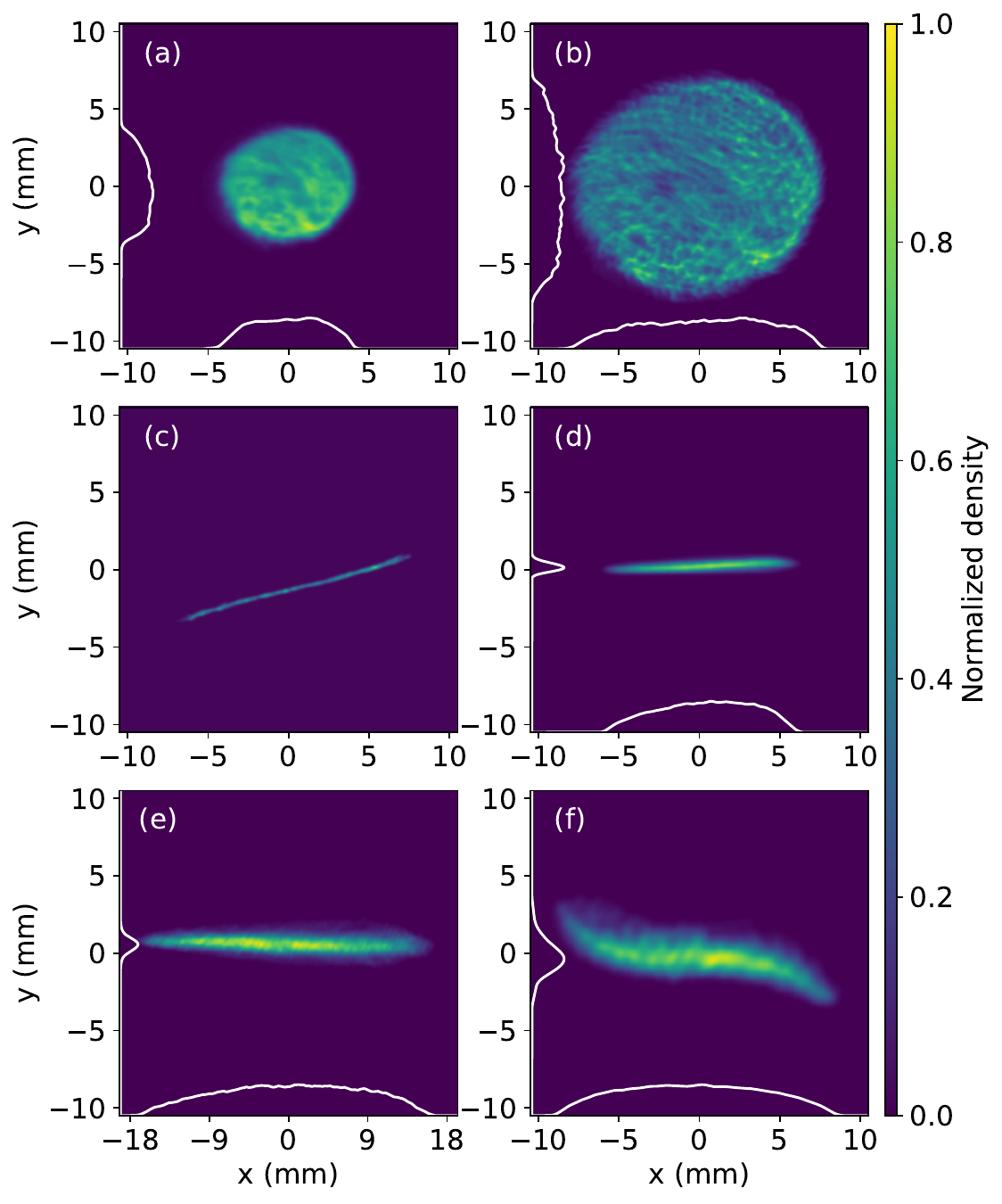}
   \caption{Measured round and flat beam distributions. 
   (a) Magnetized beam at YAG2 screen. 
   (b) Magnetized beam at YAG3 screen. 
   (c) Beamlet of magnetized beam at YAG3 screen.
   (d) Transformed flat beam at YAG2 screen.
   (e) Transformed flat beam at YAG3 screen.
   (f) Flat beam at YAG4 screen (before FRBT). }
   \label{fig:exp_RFBT_550A}
\end{figure}

\subsection{Magnetized beam generation, RFBT, and flat beam transport}

First, we generated the round, magnetized beam [see Fig.~\ref{fig:exp_RFBT_550A}(a,b)].
The high-frequency features observed in the transverse distribution are introduced by imperfections in the MLA-based homogenization of the UV laser~\cite{PhysRevAccelBeams.27.074601}.
The magnetization of the beam was inferred as~\cite{Sun:PhysRevSTAB.7.123501,PhysRevAccelBeams.22.102801}.
\begin{equation}
\Lb = \gamma\frac{\sigma_{2}\sigma_{3}\sin{\theta}}{D},
\label{eq:Lmeasurement}
\end{equation}
where $\gamma$ is relativistic factor, $\sigma_{2,3}$ are the RMS beam sizes at YAG2 and YAG3 screens shown in Fig.~\ref{fig:AWASchematic}. 
The angle $\theta$ is a rotation angle induced by the coupled motion of the magnetized beam. This angle is measured by inserting a horizontal slit at YAG2 location and measuring the slit rotation at the YAG3 position.
$D=2.84$~m is the distance between YAG2 and 3 screens.
The measured beamlet at YAG3 screen to estimate the magnetization is shown in Fig.~\ref{fig:exp_RFBT_550A}(c).
The rotation angle of the beamlet is $16.08\pm0.03$~deg, and the corresponding magnetization associated with the measured beam size and angle is $59.3\pm7.3$~\si{\micro\meter}.
If we consider the systematic error (e.g., error of UV size measurement by 10\%), then the error of the theoretical magnetization becomes $75.0\pm 14.2$~\si{\micro\meter}, where the boundary of the error covers the measured magnetization within $\pm1\sigma$.

From the measured $\Lb$, complemented by a three-screen emittance measurement using YAG$1-3$ screens we obtain and tune the Twiss parameters of the beam at the entrance of the RFBT. 
To optimize the beam parameters we employ the drive-linac solenoid magnets; see Fig.~\ref{fig:AWASchematic}.
The correlation matrix [see Eq.~\eqref{eq:Cmatrix}; where error values were computed by considering RMS beam size jitter] at the entrance of SQ1 through the three-screen diagnostics is reconstructed as
\begin{equation}
C = \left[\begin{array}{cc} 
0.15\pm0.002 & 1.79\pm0.003 \\
-0.57\pm0.001 & -0.15\pm0.003 \\
\end{array}\right],
\label{eq:correlation_matrix1}
\end{equation}
where only the averaged Twiss parameters are listed.
Using this correlation matrix, we can set the strengths of the skew triplet. 
Table~3 compares the strengths of SQ1$-$3 that are experimentally optimized with the one obtained from analytical estimate obtained using the correlation matrix described by Eq.~\eqref{eq:correlation_matrix1}. We find that the experimental settings of the quad strengths are in reasonable agreement (within 10\%) with the analytical solutions.
\begin{table}[b]
\caption{
Comparison of skew-quadrupole strengths experimentally optimized with analytical prediction for the RFBT, and measured flat beam emittances downstream of the RFBT.
}
\begin{tabular*}{\tblwidth}{@{} LRR@{} }
\toprule
\textrm{Quadrupole magnet} & \textrm{Analytic} & \textrm{Experiment} \\
\midrule
SQ1 (T/m) & $+$2.791 & $+$2.599 \\
SQ2 (T/m) & $-$3.742 & $-$3.513 \\
SQ3 (T/m) & $+$3.547 & $+$3.109 \\
\midrule
\textrm{Flat beam emittance} &  &\\
\midrule
$\epsilon_{nx}$ (\textmu{m}) & & 144.6$\pm$1.4 \\
$\epsilon_{ny}$ (\textmu{m}) & & 1.5$\pm$0.1 \\
\bottomrule
\end{tabular*}
\end{table}
\label{table:exp_skew_rfbt}

A measured flat-beam distribution downstream of the RFBT is shown in Fig.~\ref{fig:exp_RFBT_550A}(d,e). 
As expected, once the beam becomes uncorrelated against the magnetization, the flatness is well maintained even at further downstream YAG screen. 
We performed a quadrupole-scan to measured the emittance of the flat beam using DQ4 magnet and YAG2 screen.
The measured normalized emittances are $(\varepsilon_{nx}, \varepsilon_{ny})=(144.6\pm1.4, 1.5\pm 0.1)$~\textmu{m} as listed in Table~3.
The vertical emittance, is larger than the estimated value from the eigenemittance mapping. 
As reported in Ref.~\cite{PhysRevAccelBeams.25.044001}, this discrepancy between the estimation and measurement may come from the chromatic aberrations to the low emittance and machine imperfections such as misalignment of the photoinjector.

\begin{figure*}[t]
   \centering
   \includegraphics[width=\textwidth]{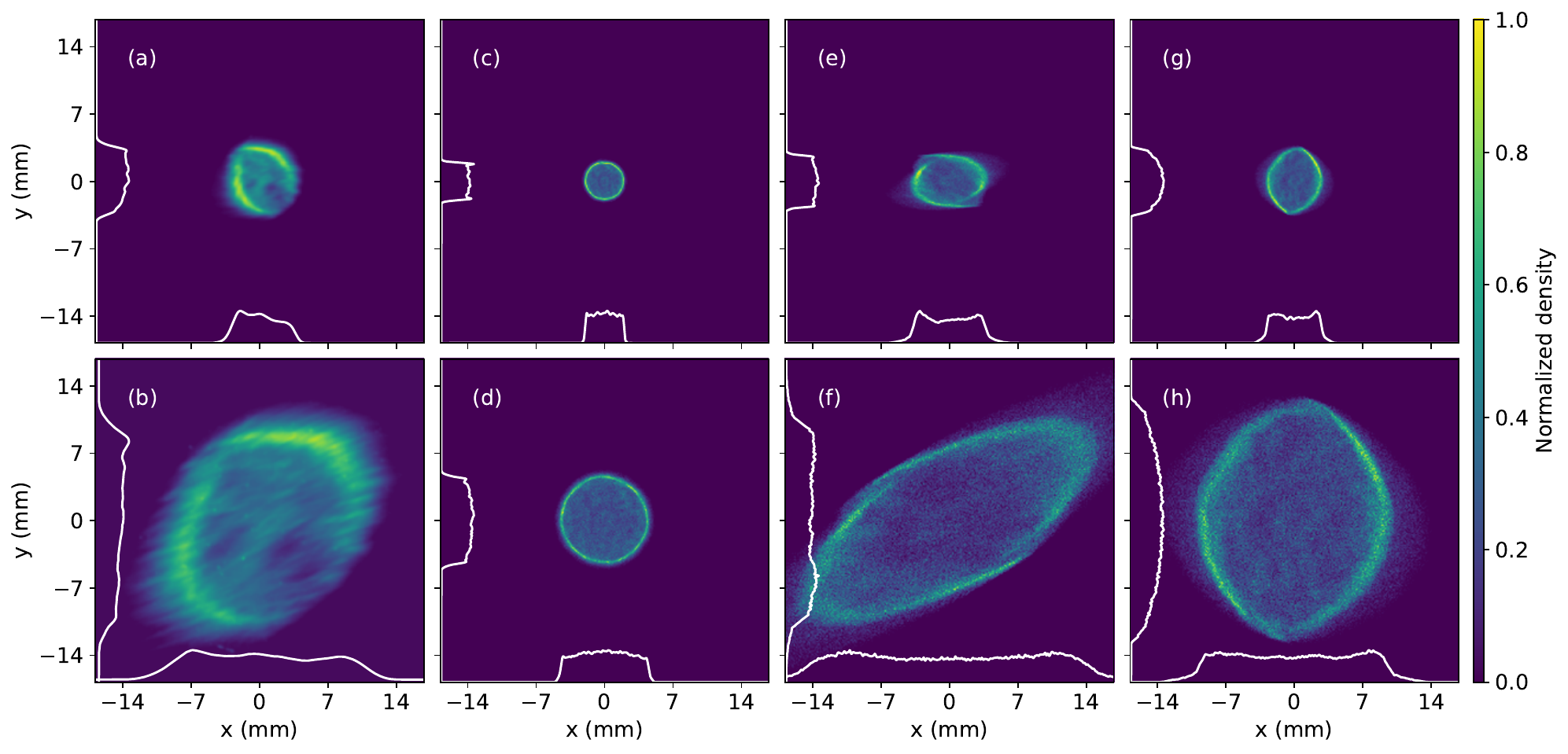}
   \caption{
   Transformed round beams at YAG5 (top row) and YAG 6 (bottom row) screens.
   (a-b) Experimental measurement.
   (c-d) Simulation case 1: fully transformed.
   (e-f) Simulation case 2: not transformed.
   (g-h) Simulation case 3: transformed.
   The skew triplet setting for each case can be seen in Table~4.}
   \label{fig:exp_frbt_550A}
\end{figure*}

The obtained flat beam  was propagated through the UTL section.
For the beam transport, we referred to the Twiss parameters obtained from the quadrupole scan using DQ4 and YAG1 screen. 
Right before the second skew-quadrupole triplet, we measured the flat beam at YAG4 screen [see Fig.~\ref{fig:exp_frbt_550A}(f)].
Here, we can clearly observe nonlinear distortion on the transverse beam distribution.
Even though the major nonlinear shape is not captured in the upstream transport beamline, we suspect that the main factor of this shape is inherent transverse space charge effect, especially from the vertical space charge from the slice charge distribution as discussed in Ref.~\cite{PhysRevAccelBeams.25.044001}.

\subsection{Flat-to-round beam transformation}

Finally, we transformed the flat beam into a round beam using second skew triplet marked as SQ$4-6$ in Fig.~\ref{fig:AWASchematic}.
During the experiment, since it was difficult to characterize the flat beam before the round transformation through such as three-screen measurement and slit-scan, we changed the method to perform the flat-to-round beam transformation; we first set the quad strengths using Eq.~\eqref{eq:FRBT_eq} with given drift space and Twiss betatron function value.
Then, we adjusted the incoming flat beam using DQ$13-15$. 
Also we used SQ4 for fine tuning of the round beam transformation.

The transformed round beam is shown in Fig.~\ref{fig:exp_frbt_550A}(a,b).
As discussed in Fig.~\ref{fig:opalsim_sigemit_beamdist}, the ring shape at the edge of the transverse profile is experimentally observed.
Furthermore, this round-beam feature is still preserved at YAG6 screen; approximately 1.38~m away from YAG5 screen.
Using YAG 5-6 screens and collimator mask, we again measured the magnetization of this transformed round beam.
The measured $\Lb$ is 74.3$\pm$ 17.7~\si{\micro\meter}, which has quantitative agreement compared to the original magnetization within 1$\sigma$.

However, since we do not clearly understand the incoming flat beam and FRBT condition, we additionally performed the tracking simulations to investigate these relations.
Here, we used OPAL-simulated flat beam at DQ4 front where the magnetization is 74.7~\si{\micro\meter}; same as that described in Section~\ref{numerical_simulations}; thus, the actual emittance and initial Twiss parameters are slightly different compared to the experimentally measured case. Nevertheless, this is to estimate the FRBT condition using the similar beam parameters.
In case of flat beam transport and flat-to-round beam transformation, we optimized the field strength of normal and skew quadrupole magnets based on the values used in the experiment.

\begin{table}[b]
\caption{
Field strength (unit of T/m) of second skew triplet used in experiment and simulations. 
In the experimental case, only the absolute values are listed. 
$\gamma\langle xy' \rangle$ and $\gamma\langle x'y \rangle$ [unit of \si{\micro\meter}] indicate transverse coupling of transformed round beam obtained by the simulations. Also $\Lb$ value of each case is shown.
}
\begin{tabular*}{\tblwidth}{@{} LRRR@{} }
\toprule
\textrm{Case} & \textrm{SQ4} & \textrm{SQ5} & \textrm{SQ6} \\
\midrule
\textrm{Experiment} & 1.473 & 3.134 & 2.795 \\
\textrm{Simulation case 1} & 1.555 & $-$2.498 & 2.958 \\
\textrm{Simulation case 2} & $-$1.381 & 1.913 & 1.948 \\
\textrm{Simulation case 3} & $-$2.204 & 4.363 & $-$10.637 \\
\midrule
\textrm{Parameters} & \textrm{$\gamma\langle xy' \rangle$} & \textrm{$\gamma\langle x'y \rangle$} & $\Lb$   \\
\midrule
\textrm{Experiment} &  &  & 74.3$\pm$17.7 \\
\textrm{Simulation case 1} & $-$74.19 & 73.96 & 74.1 \\
\textrm{Simulation case 2} & 475.00 & 337.30 & 68.8 \\
\textrm{Simulation case 3} & 81.11 & $-$62.65 & 71.9 \\
\bottomrule
\end{tabular*}
\end{table}
\label{table:frbt_skewstrength}

Figure~\ref{fig:exp_frbt_550A}(c-h) show the simulation results of transformed round beam with different skew triplet settings.
The skew setting of each case is listed in Table~4.
There are three different simulation cases.
First case, as shown in Fig.~\ref{fig:exp_frbt_550A}(c-d), is the one for complete flat-to-round beam transformation.
Here in this case, the absolute values of the skew triplet are similar to the experimental values within the error range of 13.7\%.
Also, full transformation can be seen through the transverse couplings $\gamma\langle xy' \rangle$ and $\gamma\langle x'y \rangle$ in Table~4; they are almost identical in the amount but sign is different as expected through Eq.~\eqref{eq:sigma_relation_magnetization}. 
In addition, corresponding magnetization [see Eq.~\eqref{eq:Magnetization2}] becomes the original value.
However, we can see that the envelope evolution is largely different compared to the experimental measurement.
Therefore, the settings associated with the first simulation case are not exactly matched with the experimental conditions.

In a second set of simulation we set the skew-quadrupole triplet strengths to match the experimental value within 20\%, while polarity of quadrupole focusing changed.
In this case, we can obtain similar beam distributions at each YAG screen even though the beam tilt is different compared to the experimental measurement; it can be explained by the incoming beam state and setting of the skew triplet. 
Even though the resultant $\Lb$ becomes similar to the original value within 10\% error, values of $\gamma\langle xy' \rangle$ and $\gamma\langle x'y \rangle$ are largely deviated from those of the fully transformed case in the amount of value. In addition, the sign is different compared to the completely transformed case.

In case of the final simulation, the transverse coupling is not fully symmetric in terms of absolute amount, but resultant $\Lb$ becomes original value.
At the same time, the beam shapes at each screen are similar to those found by the experiment.
Thus, it can be considered that it is round-beam-transformed case.
Nevertheless, the skew triplet setting is completely different compared to the experimental one.
This indicates that the last simulation case also does not fully represent the experimental situation.
Therefore, even though the measurements provided associated information in terms of beam distributions and magnetization value, further beam diagnostic works are needed to confirm the beam transformation on the experimental case.

\section{Conclusion}\label{conclusion}

We conducted an experiment aimed at demonstrating the ability of a cascaded round-to-flat and flat-to-round beam transformation to globally converse the beam magnetization.
The first experimental campaign reported in this paper, gives valuable insights on the process. It especially confirms that an incoming magnetized beam can be locally transformed into a flat beam and remagnetized into a round magnetized beam. However the final magnetized beam was distorted. The original of some of the distortions is understood while some would require further work.  In addition, the use of better diagnostics such as the one recently implmented in Refs.~\cite{PhysRevAccelBeams.27.074601,PhysRevLett.130.145001} using generative phase space reconstruction could provide a complete picture of the four-dimensional phase space. 

Overall, the successful, though limited, experimental results presented in this paper provide further impetus for conducting follow-up experiments with improved diagnostics.

\section{Acknowledgments}
The early concept of this research were developed in collaboration with Jefferson Lab as part of investigating magnetized electron cooling for future electron-ion collider. We acknowledge important input from Dr. Stephen Benson at Jefferson lab. This work is supported by the U.S. DOE award No. DE-AC02-06CH11357 with Argonne National Laboratory. The computing resources used for this research were provided on {\sc bebop},  a high-performance computing cluster operated by the Laboratory Computing Resource Center (LCRC) at Argonne National Laboratory. 

\printcredits

\bibliographystyle{elsarticle-num-names}


\end{document}